\documentclass[pra,aps,twocolumn,showpacs]{revtex4-1}
\usepackage{amsmath,amssymb,graphicx}
\usepackage{epsfig}
\usepackage{textcomp}
\usepackage{color}
\usepackage{epstopdf}
\usepackage{hyperref}

\hypersetup{pdfpagemode=UseNone,colorlinks=true,linktoc=page,pdfborder={0 0 0},linkcolor=blue,citecolor=blue}

\begin{document}

\title{Quantum versus classical dynamics in the optical centrifuge}
\author{Tsafrir Armon and Lazar Friedland}
\email{lazar@mail.huji.ac.il}
\doi{10.1103/PhysRevA.96.033411}

\begin{abstract}
The interplay between classical and quantum mechanical evolution in the
optical centrifuge (OC) is discussed. The analysis is based on the quantum
mechanical formalism starting from either the ground state or a thermal
ensemble. Two resonant mechanisms are identified,\ i.e. the classical
autoresonance and the quantum mechanical ladder-climbing, yielding different
dynamics and rotational excitation efficiencies. The rotating wave
approximation is used to analyze the two resonant regimes in the associated
dimensionless two-parameter space and calculate the OC excitation
efficiency. The results show good agreement between numerical simulations
and theory and are relevant to existing experimental setups.
\end{abstract}

\maketitle

\affiliation{Racah Institute of Physics, Hebrew University of Jerusalem,
Jerusalem 91904, Israel}

\affiliation{Racah Institute of Physics, Hebrew University of Jerusalem,
Jerusalem 91904, Israel}

\affiliation{Racah Institute of Physics, Hebrew University
of Jerusalem, Jerusalem 91904, Israel}

\section{Introduction}

\label{introduction} The rigid rotor is frequently used for studying the
interplay between classical and quantum mechanical phenomena, as it is
complex enough to offer intricate behavior and yet simple enough to be
conveniently handled in both theories. For example, the periodically kicked
rigid rotor problem classically may yield chaotic dynamics \cite%
{CKR1,CKR2,CKR3}, but quantum mechanically it is replaced by Anderson
localization \cite{QKR1,QKR2,QKR3}. The optical centrifuge (OC) is another
example of exploration of rotational dynamics on the molecular level.
Originally proposed and implemented by Corkum and collaborators \cite%
{Corkum1,Corkum2}, instead of a periodic drive it uses a chirped frequency
laser drive targeting the rotational degree of freedom of (mainly) diatomic
molecules. Over the last few years, several state-of-the-art experiments
\cite{Mullin,Milner1} explored the OC dynamics, demonstrating ultrafast
rotation and molecular dissociation \cite{Corkum2}, alteration of
collisional decoherence \cite{Milner2}, rotational confinement \cite{Milner3}
and even ultrafast magnetization \cite{Milner4}. The experiments use both
hot gas of light molecules, and cold gas of heavy molecules, so one could
expect to observe both quantum mechanical and classical responses.

In recent years, it was demonstrated that resonant chirped frequency drives
are very useful in studying quantum and classical phenomena in various
driven oscillatory systems, allowing exploration of the problem's phase
space. Depending on the characteristics of the system and the drive, the
evolution in these driven systems takes a classical, quantum mechanical, or
mixed form. In the classical limit, a persistent nonlinear phase locking
between the driver and the system, known as autoresonance (AR) \cite{AR},
allows for continued excitation. In contrast, in the quantum limit, the
system undergoes successive Landau-Zener (LZ) transitions \cite{Landau,Zener}%
, or quantum ladder climbing (LC). Both regimes of operation were
demonstrated and used in atoms and molecules \cite{LC1,LC1a,LC1b,LC2,AR1},
anharmonic oscillators \cite{AR}, Josephson junctions \cite{LC3}, plasma
waves \cite{PW1,PW2}, and cold neutrons \cite{Nu1}. An interesting and
surprising effect in some driven chirped anharmonic oscillators is the
forced dynamical transition from the quantum to the classical regime \cite%
{LC5}.

The optical centrifuge in its classical regime is an example of AR \cite%
{OurPaper}. But how the transition from the classical AR in this driven
system to quantum mechanical LC occurs? Of key importance in the OC is its
efficiency, i.e. the fraction of molecules excited rotationally by the
chirped laser drive. This issue was addressed recently in the AR regime of
operation \cite{OurPaper}. The corresponding quantum mechanical process was
only studied numerically \cite{OCtheory1} or under the constraint $l=m$ \cite%
{OCtheory2,OCtheory3}, where $l,m$ are the quantum numbers associated with
the total angular momentum and its projection on the laser propagation
direction, respectively. The former assumption makes it impossible to study
the response of a randomly oriented molecular ensemble to the OC pulse. The
interplay between classical and quantum mechanical effects under different
initial conditions in the system has not been studied to date. In this work,
we use the quantum mechanical description of the OC in the rigid rotor
approximation and show how it could give rise to the two different resonant
mechanisms, the AR and LC. We will find criteria separating the two regimes
in the parameter space, and calculate the corresponding OC efficiencies. We
will use numerical simulations and theory for two sets of initial
conditions, i.e. a fully populated ground-state and a "hot" thermal ensemble
and show that different combinations of parameters and initial conditions
exhibit significantly different dynamics and efficiencies.

The scope of the paper will be as follows. In Sec. \ref{Sec2}, we introduce
the model and the governing equations. Section \ref{Sec3} analyzes different
resonant regimes using the rotating wave approximation and discusses typical
excitation conditions in the associated parameter space. In the same
section, the OC efficiency under various conditions is analyzed numerically
and analytically for the aforementioned initial states. The section ends
with a discussion of the relevance of our analysis to existing experiments.
Our conclusions are summarized in Sec. \ref{summary}.

\section{The model and parameterization}

\label{Sec2} The OC uses a combination of two counter rotating and
anti-chirped circularly polarized laser beams. The resulting field has
acceleratingly rotating linear polarization, which can rotationally excite
anisotropic molecules \cite{Corkum1}. Classically, this excitation process
is an example of rotational autoresonance \cite{OurPaper}, as the molecule
continuously self-adjusts its rotation frequency to that of the accelerating
rotation of drive. For a driving wave propagating along the $Z$ axis, with
polarization angle $\phi _{d}\left( t\right) $ in the $XY$ plane, after
averaging over the optical frequency of the laser beams, the interaction
potential energy of a diatomic molecule in spherical coordinates is given by
$U=-\varepsilon \sin ^{2}\theta \cos ^{2}\left( \varphi -\phi _{d}\right) $
\cite{Corkum1}, where $\varepsilon =\left( \alpha _{\parallel }-\alpha
_{\perp }\right) E_{0}^{2}/4$, $\alpha _{\parallel }$,$\alpha _{\perp }$ are
the polarizability components of the molecule and $E_{0}$ is the electric
field amplitude of the combined laser beam. Similarly to existing
experimental systems \cite{Milner1,Milner2,Milner3}, we will use a drive
with zero initial frequency and linear frequency chirp $\omega _{d}=d\phi
_{d}/dt=\beta t/2$, where $\beta >0$ is the chirp rate.

We will analyze the OC dynamics governed by the full quantum mechanical
Hamiltonian $\widehat{H}=\widehat{H}_{0}+\widehat{U}$, using the set of
eigenstates of the unperturbed Hamiltonian $\hat{H_{0}}=\hat{L}^{2}/2I$,
where $\hat{L}$ is the angular momentum operator, and $I$ is the molecule's
moment of inertia. This set comprises the spherical harmonics $\left\vert
l,m\right\rangle $, satisfying $\hat{L}^{2}\left\vert l,m\right\rangle
=\hbar ^{2}l\left( l+1\right) \left\vert l,m\right\rangle $, $\hat{L_{z}}%
\left\vert l,m\right\rangle =\hbar m\left\vert l,m\right\rangle $, where $%
\hat{L_{z}}$ is the operator associated with the projection of angular
momentum on the $Z$ axis \cite{Sakurai}. At this stage, we can identify
three relevant time scales, i.e. the drive sweeping time $t_{s}=1/\sqrt{%
\beta }$, the Rabi (driving) time scale $t_{d}=\hbar /\varepsilon $ and the
characteristic quantum mechanical "rotation" time $t_{c}=I/\hbar $. The
three time scales yield two dimensionless parameters:
\begin{equation}
P_{1}=\frac{t_{s}}{t_{d}}=\frac{\varepsilon }{\hbar \sqrt{\beta }},
\label{P1}
\end{equation}%
and
\begin{equation}
P_{2}=\frac{t_{s}}{t_{c}}=\frac{\hbar }{I\sqrt{\beta }},  \label{P2}
\end{equation}%
characterizing the driver's strength and the problem's nonlinearity,
respectively. This parameterization yields the classical
parametrization \cite{OurPaper}, if one replaces the quantum mechanical
action scale $\hbar $ by the action scale $\sqrt{Ik_{B}T}$ of a thermal
classical ensemble, where $k_{B}$ is the Boltzmann constant and $T$ is the
temperature.

The form of the interaction $\hat{U}$ leads to selection rules, where
transitions are allowed to states with $\Delta l,\Delta m$ equal to $0$ or $%
\pm 2$ only. This also follows from the two-photon nature of these Raman
processes and guarantees the conservation of parity for both $l,m$.
Expressing the wave function in the Schrodinger equation for the driven
problem as $\sum_{l,m}a_{l,m}\left\vert l,m\right\rangle $, the
dimensionless evolution equation for coefficient $a_{l,m}$ in terms of
parameters $P_{1,2}$ is given by:
\begin{equation}
i\frac{da_{l,m}}{d\tau }=E_{l}a_{l,m}+P_{1}\sum_{\substack{ \Delta l=  \\ %
0,\pm 2}}\sum_{\substack{ \Delta m=  \\ 0,\pm 2}}c_{l,m}^{\Delta l,\Delta
m}a_{l^{\prime },m^{\prime }}e^{i\Delta m\phi _{d}},  \label{NRWA_SE}
\end{equation}%
where the time derivative is with respect to the slow dimensionless time $%
\tau =\sqrt{\beta }t$, $l^{\prime }=l+\Delta l$, $m^{\prime }=m+\Delta m$
and $E_{l}={P_{2}}l\left( l+1\right) /2$. The details of the derivation, and
the the coupling coefficients $c_{l,m}^{\Delta l,\Delta m}$ are given in
Appendix \ref{AppA}.

The evolution described by Eq. (\ref{NRWA_SE}) exhibits different dynamics
depending on parameters $P_{1,2}$ and initial condition. In this paper, we
focus on two types of initial conditions, i.e. a fully populated ground
state ($l=0$) and a finite temperature thermal state. For the purpose of
this work, it is convenient to define the temperature via the characteristic
$l$ value, $l_{c}\ge 0$, given by equating the thermal and rotational
energies:
\begin{equation}
k_{B}T=\frac{\hbar ^{2}}{2I}l_{c}\left( l_{c}+1\right) .
\end{equation}

\begin{figure}[tbp]
\includegraphics[width=3.375in]{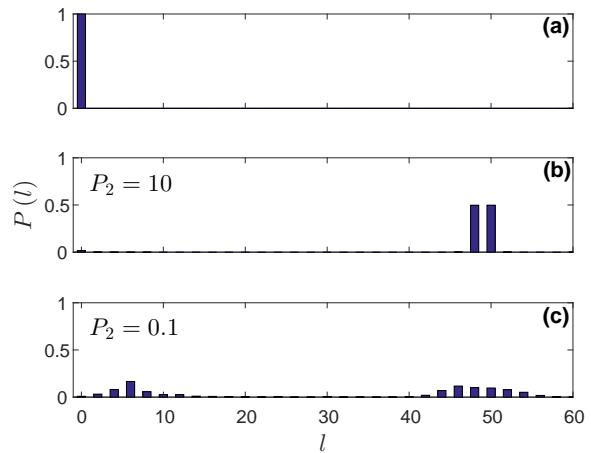}
\caption{The normalized distribution of the $l$-states from the numerical
solution of Eq. (\protect\ref{NRWA_SE}) for the ground state initial
condition shown in panel (a). The resulting final distributions for $%
P_{2}=10 $ are shown in panel (b) and for $P_{2}=0.1$ in panel (c). The
parameter $P_{1}=10$ in both panels (b) and (c) and the final time is $%
\protect\tau=99P_{2}$ .}
\label{Fig1a}
\end{figure}

\begin{figure}[tbp]
\includegraphics[width=3.375in]{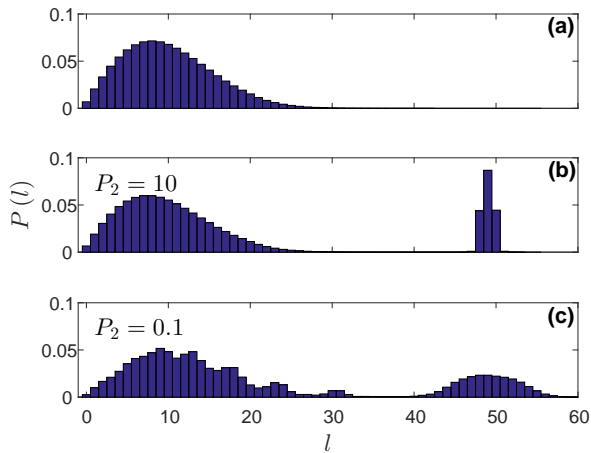}
\caption{The normalized distribution of the $l$-states from the numerical
solution of Eq. (\protect\ref{NRWA_SE}) for a thermal initial condition with
$l_{c}=11.5$ (a). The resulting final distributions for $P_{2}=10$ are shown
in panel (b) and for $P_{2}=0.1$ in panel (c). The parameter $P_{1}=10$ in
both panels (b) and (c) and the final time is $\protect\tau =99P_{2}$.}
\label{Fig1b}
\end{figure}
Figures \ref{Fig1a} and \ref{Fig1b} present numerical solutions of Eq. (\ref%
{NRWA_SE}) starting from the ground state ($l=0$) (Fig. \ref{Fig1a}) and a
thermal ensemble with $l_{c}=11.5$ (Fig. \ref{Fig1b}). These initial
conditions ($\tau =0$) are shown on the top panels (a) in the figures, while
the other panels show the final state at $\tau =99P_{2}$ for $P_{1}=10$. The
values of $P_{2}$ were $10$ (panel b) and $0.1$ (panel c). The final driving
frequency matches the resonant transition $l=48\rightarrow l=50$ (as will be
explained below) and the resonantly excited population around this target
state illustrates the results of different dynamics. Indeed, due to the
conservation of parity of both $l,m$ when starting in the ground state, only
even $l$ levels are excited, in contrast to the thermal ensemble where both
even and odd excited states are present. Furthermore, for both initial
conditions, the width of the population around the target state decreases
dramatically with $P_{2}$. Finally, the fraction of excited population
around the target state ranges from as high as $\sim 100\%$ (Fig. \ref{Fig1a}%
b) to as low as $\sim 17\%$ (Fig. \ref{Fig1b}b). We proceed to studying
these characteristic evolutions next.


\section{Rotational LC versus classical AR}

\label{Sec3}

\subsection{Resonant Evolution}

\label{SubSecA} In studying the different responses of the system to the
chirped frequency drive, we examine the resonant interactions, which give
rise to both the quantum mechanical LC and classical AR. The interaction $%
\hat{U}$ yields the coupling of each state $\left\vert l,m\right\rangle $ to
itself and, in general, $8$ other states. However, not all of these
transitions are resonant, and, to proceed, we apply the rotating wave
approximation (RWA), the validity of which will be discussed below. We
identify the nearest resonant transition $\left\vert l,m\right\rangle
\rightarrow \left\vert l+2,m+2\right\rangle $ and apply the RWA, neglecting
all nonresonant terms. This resonant transition conserves the difference $%
C=l-m$ and, therefore, the $m$ index is omitted in the following equations
describing a given $C$ value. By transforming Eq. (\ref{NRWA_SE}) to the
rotating frame of reference, i.e defining $W_{l,m}=e^{il\phi _{d}}a_{l,m}$,
and neglecting all nonresonant (rapidly oscillating) driving terms we get

\begin{equation}
i\dot{W_{l}}=\Gamma _{l}W_{l}+P_{1}c_{l}^{2}W_{l+2}+P_{1}c_{l}^{-2}W_{l-2},
\label{RWA_SE}
\end{equation}%
where $\Gamma _{l}=E_{l}+P_{1}c_{l}^{0}-l\tau /2$. The coefficient $%
c_{l}^{0} $, which represents some energy shift, does not vary significantly
between the coupled states, and its contribution in $\Gamma $ can usually be
ignored. Then, the coupling matrix of a single Landau-Zener type \cite%
{Landau,Zener} two-level transition $l-2\rightarrow l$ is
\begin{equation}
\left(
\begin{array}{cc}
\frac{P_{2}}{2}\left( l-2\right) \left( l-1\right) -\frac{\tau }{2}\left(
l-2\right) & P_{1}B_{l} \\
P_{1}B_{l} & \frac{P_{2}}{2}l\left( l+1\right) -\frac{\tau l}{2}%
\end{array}%
\right) ,  \label{2LevelMat}
\end{equation}%
where $B_{l,m}=c_{l-2,m-2}^{2,2}=c_{l,m}^{-2,-2}$ and again index $m$ is
omitted. Following the reasoning of Ref. \cite{LC2,LC5}, for having
successive Landau-Zener (LZ)\ transitions in our chirped system, the
duration of each transition must be much shorter than the time between two
successive transitions. The time $\tau _{l}$ of each transition is given by
the energy crossing condition, or $\Delta \Gamma =\Gamma _{l}-\Gamma
_{l-2}=0 $, yielding $\tau _{l}=P_{2}\left( 2l-1\right) $, so the time
between successive transitions is $4P_{2}$. The duration of a transition is $%
O\left( 1\right) $ when $\left\vert B_{l}P_{1}\right\vert $ is small and $%
O\left( \left\vert B_{l}P_{1}\right\vert \right) $ when it is large \cite%
{LC5}. Therefore, we estimate the duration of the transition as $%
1+\left\vert B_{l}P_{1}\right\vert $. Consequently, the condition for the
successive LC process is:
\begin{equation}
P_{2}\gg \frac{1}{4}+\frac{P_{1}}{16},  \label{QCond}
\end{equation}%
where we took $\left\vert B_{l}\right\vert $ at its maximal value of $1/4$
for all $C$ (see Appendix \ref{AppA}).

When condition (\ref{QCond}) is met, the transitions are well separated,
only two states are coupled at a time, and LC takes place. When
transitioning to an unpopulated state, The transition probability in a
single LZ step is given by \cite{Landau,Zener}:
\begin{equation}
P_{l-2\rightarrow l}=1-\exp [-2\pi (P_{1}B_{l})^{2}].  \label{LZformula}
\end{equation}%
\ The efficiency of this process is governed by $P_{1}$ only, and when its
value is sufficiently large the transitions could yield nearly $100\%$
population transfer. If the OC proceeds from the ground state and the
chirped driving frequency passes the resonance with some higher state $%
\widehat{l}$, the fraction of rotationally excited population with $l\geq
\widehat{l}$ will be
\begin{equation}
f(\widehat{l})=\prod_{n=1}^{\widehat{l}/2}\left\{ 1-\exp \left[ -2\pi
(P_{1}B_{2n})^{2}\right] \right\} .  \label{F}
\end{equation}%
However, if condition (\ref{QCond}) is not met, several states are coupled
simultaneously, the interaction becomes increasingly classical, and the
classical AR may take place. This classical version of the OC was discussed
in \cite{OurPaper} and one expects the correspondence principle to hold in
the $l,m\gg 1$ limit. In particular, the classical single resonance
approximation in the AR theory yields the same conservation law as with the
RWA, i.e. $L-L_{z}=const$. Furthermore, in the $l,m\gg 1$, the resonant
coupling coefficients become%
\begin{eqnarray*}
c_{l,m}^{0,0} &\Longrightarrow &-\frac{1}{2}+\frac{1}{4}\left( 1-\frac{m^{2}%
}{l^{2}}\right) , \\
B_{l,m} &\Longrightarrow &-\frac{1}{16}\left( 1+\frac{m}{l}\right) ^{2},
\end{eqnarray*}%
which, using the semiclassical approximation $L\approx \hbar l$, coincides
with the classical interaction functions $F(L_{z}/L),V(L_{z}/L)/2$ in Eqs.
(6) and (7) in \cite{OurPaper,SideNote1}. The classical analysis also shows
that the capture into rotational autoresonance is possible only if
\begin{equation}
P_{1}^{cl}P_{2}^{cl}>1/2,  \label{ClassicalThreshold}
\end{equation}%
where the classical dimensionless parameters are\ $P_{1}^{cl}=\varepsilon /%
\sqrt{Ik_{B}T\beta }$ and $P_{2}^{cl}=\sqrt{k_{B}T/I\beta }$. This result
has its correspondence in the quantum problem as well, because $P_{1}P_{2}=$
$P_{1}^{cl}P_{2}^{cl}$. It should be noted that other nonlinear oscillators
studied in this context exhibited a dynamical transition from LC to AR as a
result of an unbounded growth of the coupling coefficient (here, $B_{l}$)
\cite{LC5}. In the present case, this coefficient does grow (in absolute
value), but its growth is bounded, preventing a dynamical transition between
the two regimes.

Finally, we discuss the validity of the RWA in our problem. This
approximation is valid if the dimensionless frequencies of the non-resonant
terms neglected in Eq. (\ref{RWA_SE}) are large enough. One can show that
for a given $l$, where all nine allowed transitions exist, $P_{2}\left(
2l-1\right) $ is the smallest of these frequencies. Then, by estimating the
duration $\Delta \tau $ of a typical resonant transition as being of $O(1)$,
the inequality $P_{2}\left( 2l-1\right) \gg 1$ must hold for the validity of
RWA.  For thermal ensembles with $l_{c}\gg \left( 1+P_{2}\right) /2P_{2}$
the overall RWA validity remains good, since the population of $l$ states
violating RWA in such ensembles is relatively small.

\subsection{Ground-state versus thermal initial condition}

\begin{figure*}[t]
\includegraphics[width=6.75in]{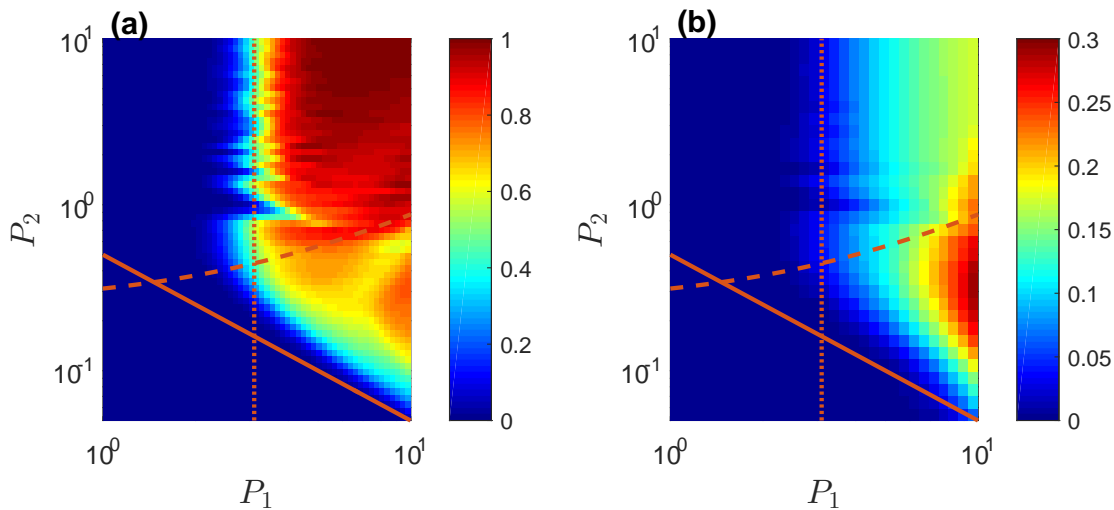} .
\caption{Color coded excitation efficiency from simulations in $P_{1,2}$
parameter space. The left panel represents simulations starting in the
ground state without RWA, while the right panel shows the results for a
thermal initial condition with $l_{c}=11.5$ and using the RWA. The final
chirp time is $\protect\tau_{f}=99P_{2}$ , which corresponds to $l_{f}=50$.
The different lines show the quantum-classical boundary (dashed line),
efficient LC threshold (dotted line) and AR boundary (solid
line).}
\label{Fig2}
\end{figure*}

Here we discuss the efficiency of rotational excitation in the OC under two
distinct initial conditions as illustrated in Figures \ref{Fig1a} and \ref%
{Fig1b}, i.e. starting either in the ground state ($l_{c}=0$) or a "hot"
thermal ensemble ($l_{c}=11.5$), respectively. The latter value of $l_{c}$
is characteristic of existing experiments, such as in $N_{2}$ or $O_{2}$ at
room temperature \cite{Milner1}. We define the excitation efficiency as the
fraction of rotationally excited molecules within $20\%$ from the final
target state $l_{f}=\frac{1}{2}+\frac{\tau _{f}}{2P_{2}}$, $\tau _{f}$ being
the final driving time. Figures \ref{Fig2}a and \ref{Fig2}b show numerically
found excitation efficiency for the ground-state initial condition and for
the thermal initial ensemble, respectively (note that the color scales in
the two figures are different). The final driving time in these examples is $%
\tau _{f}=99P_{2}$ corresponding to the resonant transition $l=48\rightarrow
l=50$ ($l_{f}=50)$, so we used the fraction of the molecules excited beyond $%
\widehat{l}=0.8l_{f}=40$ in defining the excitation efficiency $f$. We show
the quantum/classical separation boundary (\ref{QCond}) (dashed line),
as well as the autoresonance boundary line (\ref{ClassicalThreshold}) (solid line) in both figures bounding the domains of different resonant excitation
mechanisms. The value of $P_{1}\approx 3.1$ for which $f=0.5$ according to
Eq. (\ref{F}) can serve as the threshold for high excitation efficiency in
the LC regime. We show this value of $P_{1}$ in Fig. \ref{Fig2} by the
vertical dotted lines.

\begin{figure}[tbp]
\includegraphics[width=3.375in]{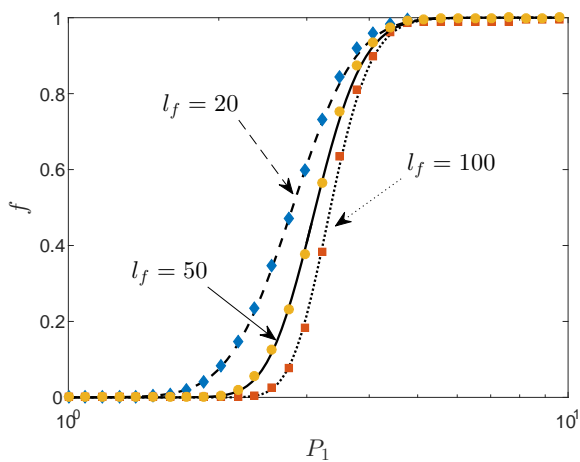}.
\caption{The OC excitation efficiency from numerical simulations for the
ground-state initial condition (markers) and the theoretical prediction, Eq.
(\protect\ref{F}) (lines). The parameter $P_{2}=10$ is kept constant, while
the final $l_{f}$=$20$ (blue diamonds, dashed line), $50$ (orange circles,
solid line), and $100$ (red squares, dotted line).}
\label{Fig3}
\end{figure}

We discuss the ground state initial condition (Fig. \ref{Fig2}a) first. Note
that in the quantum region in this case (above the quantum/classical
separation line), the resonant excitation efficiency increases with $P_{1}$,
is almost independent of $P_{2}$, and can reach nearly $100\%$. This can be
explained via the LC arguments. Indeed, for calculating the efficiency $f$
in this case one uses Eq. (\ref{F}) with $C=0$. Figure \ref{Fig3} compares
the prediction of Eq. (\ref{F}) with simulations for $P_{2}=10$ (quantum
regime) and three values of the final driving time defined by the target
states $l_{f}=20$ (blue diamonds, dashed line), $50$ (orange circles, solid
line), and $100$ (red squares, dotted line). All cases show good agreement
between theory and simulations, demonstrating the validity of the RWA, and
the possibility of very high (nearly $100\%$) excitation efficiencies for
sufficiently large $P_{1}$. Nonetheless, for a given $P_{1}$, increasing the
target state $l_{f}$ reduces the excitation efficiency because more
population is left behind as the number of the successive LZ steps grows.
This situation differs significantly from the classical AR, where molecules
trapped in the rotational resonance are not lost and, in principle, can
increase their rotational energy indefinitely as the laser pulse chirp
continues, until other effects become important. If one starts from the
ground state in the classical domain of parameters $P_{1,2\text{ }}$in \ref%
{Fig2}a, the RWA is not valid, and nonresonant transitions, which break the
conservation of $C=l-m$ play a key role in the removal of population from
the resonant pathway. Nevertheless, the autoresonant boundary line serves as
a threshold for efficient rotational excitation even for this initial
condition despite its initial quantum nature. This case is illustrated in a
video (See Supplemental Material \cite{Movies}) showing the evolution of
rotational population in the $l,m$ space, for parameters identical to those
of Fig. \ref{Fig1a}c.

\begin{figure}[tbp]
\includegraphics[width=3.375in]{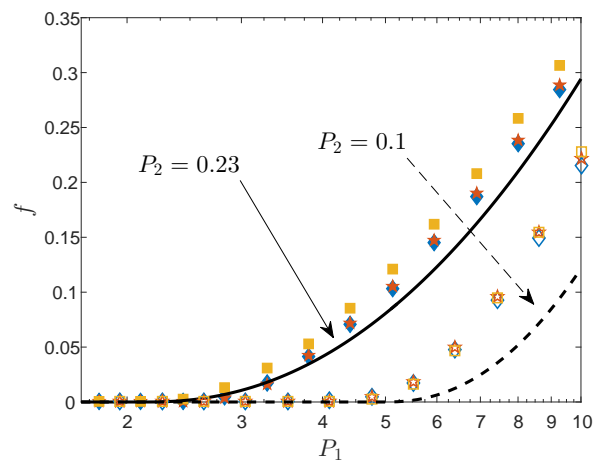}
\caption{The excitation efficiency for a thermal ensemble from numerical
simulations with all parameters identical to those in in Fig. \protect\ref%
{Fig2}, but $P_{2}=0.23$ (filled markers) and $0.1$ (empty markers). The
markers show three different simulations: with RWA (blue diamonds), without
RWA (red pentagram), and the classical Monte-Carlo simulation (orange
squares), while the lines are theoretical predictions from \protect\cite%
{OurPaper}.}
\label{Fig4}
\end{figure}

In the case of a thermal initial state, the classical region of the
parameter space exhibits the highest resonant excitation efficiency. The
latter is described by the classical theory of Ref. \cite{OurPaper} (see Eq.
(28) in that paper). This theory uses an additional weak drive assumption,
which in terms of the parameters of the present work can be written as $%
\sqrt{\frac{2}{l_{c}\left( l_{c}+1\right) }\frac{P_{1}}{P_{2}}}\ll 1$ (note
that $\hbar $ cancels out in this expression). Figure \ref{Fig4} compares
the OC excitation efficiency in simulations using quantum mechanical
formalism with (blue diamonds) and without (red pentagrams) RWA with the
predictions of the classical theory (lines) and Monte Carlo simulations
(orange squares). We used $l_{c}=11.5$, and $P_{2}=0.23$ (filled markers,
solid line) and $0.1$ (empty markers, dashed line) in these calculations.
The agreement with the classical theory of \cite{OurPaper} is quite good for
$P_{2}=0.23$, but not as good for $P_{2}=0.1$, because of the breaking of
the weak drive assumption. Nevertheless, the classical Monte-Carlo
simulations (orange squares) show good agreement with the quantum
simulations even for $P_{2}=0.23$, which is close to the quantum/classical
separation line. As expected, when the value of $P_{2}$ is decreased, the
agreement between the simulations gets better. Note that while the RWA is
not strictly valid in the region of the parameter space in these
simulations, the results with and without RWA show good agreement due to the
considerations described at the end of subsection \ref{SubSecA}.
Consequently, we have used the RWA in simulations in Fig. \ref{Fig2}b,
allowing a significant reduction of numerical complexity of quantum
simulations (see appendix \ref{AppB}). To the best of our knowledge, there
is no analytic theory for calculating the excitation efficiency in the
quantum region for initially thermal ensembles. Nevertheless, the evolution
in this regime has the characteristics of LC, as exemplified in Fig. \ref%
{Fig1b}b and in a movie (See Supplemental Material \cite{Movies}) showing
the evolution for the same parameters in the $l,m$ space. The successive LC
transitions still take place, but now there exists a width in $C=l-m$. Note
that the width in $l$ as seen in the simulations of the moving resonant
bunch, is actually two different resonant pathways experiencing LC, each
representing the conserved parity of $l$.

\subsection{Relevance to existing experiments}

\begin{figure}[tbp]
\includegraphics[width=3.375in]{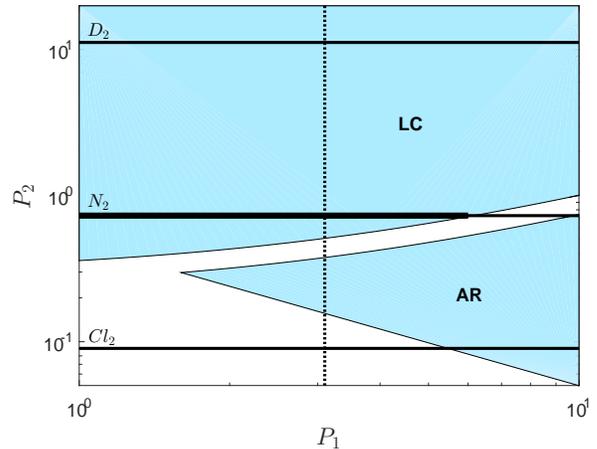}
\caption{The $P_{1,2}$ parameter space with shaded areas dividing it
into AR and LC regimes. The value of $P_{2}$ for $\protect\beta%
=1ps^{-2}$ and $D_{2}$, $N_{2}$ and $Cl_{2}$ molecules is represented by the horizontal solid lines. The dotted line shows the
efficient LC threshold, while the thick part of the $N_{2}$ line represents
the trajectory of the simulationss in Fig. \ref{Fig6}.}
\label{Fig5}
\end{figure}

Lastly, it is important to discuss our analysis in the context of existing
experimental setups. Characteristic value of the chirp rate $\beta $ in
these setups is $1ps^{-2}$ (see, for example, \cite{beta1,beta2}). With this
value of\ $\beta $, and the molecules already in use in OC experiments \cite%
{Corkum2,ExpRes,beta2},
parameter $P_{2}$ (see Eq. \ref{P2}) varies from $11.2$ (for $Cl_{2}$) through $0.73$ ($N_{2}$) to $0.09$ ($D_{2}$). These three values are represented by horizontal lines in Fig \ref{Fig5} in the $P_{1,2}$ parameter space. In the same figure, we also show the AR
and LC domains (shaded blue areas) as discussed above. Clearly, these
different resonant regimes are accessible in experiments. One can also see
that in the $D_{2}$ and  $N_{2}$ cases, one can exploit both the \ AR and LC by
a proper choice of $P_{1}$, while $Cl_{2}$ can not exhibit quantum LC dynamics.

\begin{figure}[tbp]
\includegraphics[width=3.375in]{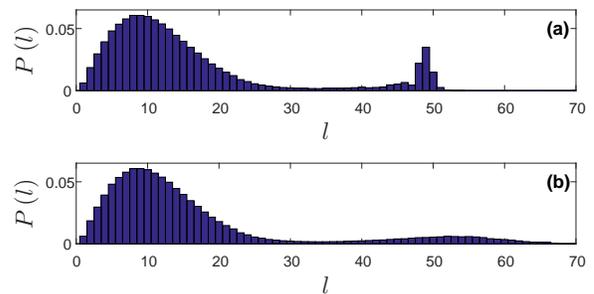}
\caption{The distributions $P\left(l\right)$ in simulations with truncation of the laser pulse (panel a), and without the truncation (panel b).}
\label{Fig6}
\end{figure}

To further exploit our analysis, we address the experimental results of Ref.
\cite{ExpRes}. The experiment involved $N_{2}$  molecules ($P_{2}=0.73$) and
the OC laser pulse had a varying amplitude of a Gaussian form, $%
P_{1}=P_{10}\exp (-\tau ^{2}/2\sigma ^{2})$, $\sigma =52$ with $\tau $
starting from zero \cite{Valery}. The lower panel in Fig. 10.8 of \cite%
{ExpRes} shows two results with very different $l-$width of the excited
bunch of molecules,  with the narrow bunch corresponding to the laser pulse
truncated at $\tau \approx 97$. Our analysis suggests the following
interpretation of these results. If initially the system evolves in the
efficient LC regime, the excited bunch is narrow (2-3 excited $l$ states).
As parameter $P_{1}$ decreases, one crosses the efficient LC excitation
threshold (vertical dotted line in Fig. \ref{Fig5}) at some time and, as a
result,  more and more population leaves the resonant bunch and stays
behind, until the bunch vanishes completely. This leads to a broad (non
resonant)\ excited population. However, if one truncates the laser pulse at
earlier time, the population freezes and the bunch remains narrow. To check
this hypothesis, we have used our simulations and present the results in
Fig. \ref{Fig6}. The upper and the lower panels in the figure show the distribution $P\left( l\right)$ with and without the truncation, respectively. In this simulation   $P_{10}=6$ and the
system evolves in the $P_{1,2}$ parameter space along the thick part of the $%
N_{2}$ line in Fig. \ref{Fig5}. One can observe formation of either narrow
or wide excited bunches similar to the experimental results \cite{ExpRes}.

\section{SUMMARY}

\label{summary} In conclusion, we have studied the problem of resonant
rotational excitation in the optical centrifuge for a wide range of
parameters starting from either the ground-state or a "hot" thermal
ensemble. Based on three characteristic time scales in the problem, we
introduced two dimensionless parameters $P_{1,2}$, and studied the resonant
nature of the problem in the parameters space by using the rotating wave
approximation. We have shown how two distinct resonant regimes can appear in
this problem, i.e the quantum mechanical ladder-climbing and the classical
autoresonance and discussed a separation criterion between the two regimes
in the $P_{1,2}$ parameter space.

We have also derived criteria for efficient rotational excitation in the OC
for the two resonant mechanisms and have shown that both are present with
the aforementioned initial conditions, but their manifestation is different.
Indeed, the maximal resonant excitation efficiency is significantly higher
with the ground-state initial condition. Furthermore, the most efficient
excitation mechanism is the ladder climbing in the case of the ground-state
initial condition, while it's the classical autoresonance when starting with
the thermal ensemble. When possible, the excitation efficiency in
simulations was compared to theoretical predictions. Our current theoretical
understanding allows calculation of the excitation efficiency in the most
efficient regime for each of the above initial conditions. The validity of
the rotating wave approximation and quantum/classical correspondence was
also studied analytically and numerically.

The results of this work combine the classical \cite{OurPaper} and the
quantum formalisms, broadening the previous analysis of the OC problem,
which did not address the full complexity of the quantum case, especially
dealing with thermal initial conditions \cite{OCtheory3}. These results
address main issues associated with the efficiency and the spectral width of
the excited resonant bunch of molecules in the OC, which is important in
planning future experiments. We have also shown that existing experimental
setups can access different resonant regimes of operation studied in this
work by using light and heavy molecules, varying gas temperature, and laser intensity. While the
analysis presented here assumes rigid rotor molecules, some effects of
nonrigidity could be studied similarly in the future. For example, the
limitation on parity of $l,m$ due to spin-statistics in some molecules does
not change the analysis. On the other hand, the centrifugal radial expansion
adds a third parameter to the problem, representing the $\propto l^{2}\left(
l+1\right) ^{2}$ addition to the energy. Since this effect continuously
increases the time between the resonant transitions, it may allow a forced
dynamical transition from classical autoresonance to quantum ladder-climbing
in the process of the same continuing rotational excitation.

\begin{acknowledgments}
This work was supported by the Israel Science Foundation grant 30/14.
\end{acknowledgments}


\appendix

\section{Quantum mechanical coupling}

\label{AppA} Using the parameterization of Sec. \ref{Sec2}, the
dimensionless evolution equation for the coefficients $a_{l,m}$ is:
\begin{equation}
i\frac{da_{l,m}}{d\tau }=E_{l}a_{l,m}+\left\langle l,m\right\vert \hat{U}%
\sum_{k,s}a_{k,s}\left\vert k,s\right\rangle .  \label{App1Eq1}
\end{equation}%
Here we expand the dimensionless interaction energy $U$ in spherical
harmonics $Y_{l}^{m}\left( \theta ,\phi \right) $:
\begin{equation}
U=-P_{1}\left[ \sqrt{\frac{2\pi }{15}}\left( Y_{2}^{2}e^{-2i\phi
_{d}}+Y_{2}^{-2}e^{2i\phi _{d}}\right) -\sqrt{\frac{4\pi }{45}}Y_{2}^{0}+%
\frac{1}{3}\right] .
\end{equation}%
Then, the inner product in Eq. (\ref{App1Eq1}) can be expressed as the
integral of three spherical harmonics, and represented via the Wigner 3-j
symbol. The selection rules for the quantum mechanical transitions occur
naturally from the selection rules of the 3-j symbol, while the coupling
coefficients $c_{l,m}^{\Delta l,\Delta m}$ can be calculated directly. We
summarize these coefficients (up to the phase term which was included
explicitly in Eq. (\ref{NRWA_SE})) in table \ref{table1}.

\begin{table}[b]
\caption{Coupling coefficient for the transition $\left|l,m\right>%
\rightarrow \left|l+\Delta l, m+\Delta m\right>$}
\label{table1}%
\begin{tabular}{|c|c|c|}
\hline
$\Delta l$ & $\Delta m$ & $c^{\Delta l,\Delta m}_{l,m}$ \\ \hline
$0$ & $0$ & $-\frac{1}{3}\left[1-\frac{l^2+l-3m^2}{\left(2l-1\right)%
\left(2l+3\right)}\right]$ \\ \hline
$2$ & $0$ & $\frac{1}{2}\sqrt{\frac{\left(l-m+1\right)\left(l-m+2\right)%
\left(l+m+1\right)\left(l+m+2\right)}{\left(2l+1\right)\left(2l+3\right)^2%
\left(2l+5\right)}}$ \\ \hline
$-2$ & $0$ & $\frac{1}{2}\sqrt{\frac{\left(l-m-1\right)\left(l-m\right)%
\left(l+m-1\right)\left(l+m\right)}{\left(2l+1\right)\left(2l-1\right)^2%
\left(2l-3\right)}}$ \\ \hline
$0$ & $2$ & $\frac{1}{2}\frac{\sqrt{\left(l+m+1\right)\left(l+m+2\right)%
\left(l-m-1\right)\left(l-m\right)}}{\left(2l-1\right)\left(2l+3\right)}$ \\
\hline
$0$ & $-2$ & $\frac{1}{2}\frac{\sqrt{\left(l-m+1\right)\left(l-m+2\right)%
\left(l+m-1\right)\left(l+m\right)}}{\left(2l-1\right)\left(2l+3\right)}$ \\
\hline
$2$ & $2$ & $-\frac{1}{4}\sqrt{\frac{\left(l+m+1\right)\left(l+m+2\right)%
\left(l+m+3\right)\left(l+m+4\right)}{\left(2l+1\right)\left(2l+3\right)^2%
\left(2l+5\right)}}$ \\ \hline
$2$ & $-2$ & $-\frac{1}{4}\sqrt{\frac{\left(l-m+1\right)\left(l-m+2\right)%
\left(l-m+3\right)\left(l-m+4\right)}{\left(2l+1\right)\left(2l+3\right)^2%
\left(2l+5\right)}}$ \\ \hline
$-2$ & $2$ & $-\frac{1}{4}\sqrt{\frac{\left(l-m\right)\left(l-m-1\right)%
\left(l-m-2\right)\left(l-m-3\right)}{\left(2l+1\right)\left(2l-1\right)^2%
\left(2l-3\right)}}$ \\ \hline
$-2$ & $-2$ & $-\frac{1}{4}\sqrt{\frac{\left(l+m\right)\left(l+m-1\right)%
\left(l+m-2\right)\left(l+m-3\right)}{\left(2l+1\right)\left(2l-1\right)^2%
\left(2l-3\right)}}$ \\ \hline
\end{tabular}%
\end{table}

\section{Numerical simulations}

\label{AppB} Our numerical simulations when starting in the ground state
used Eq. (\ref{NRWA_SE}). Because of the preferred resonant transition $%
\left\vert l,m\right\rangle \rightarrow \left\vert l+2,m+2\right\rangle $,
even for large time intervals, the value of $C=l-m$ remained bounded
throughout the evolution (even when the RWA fails initially). Therefore, for
faster simulations, a maximum value $C_{max}$ was chosen and only states
with $C\leq C_{max}$ were taken into account. Furthermore, due to the parity
conservation of $l,m$, only states with even $l,m$ were considered.

For the thermal state initial condition, the von Neumann equation was solved
\begin{equation}
i\frac{d\rho }{d\tau }=[H,\rho ],
\end{equation}%
where $\rho $ is the density matrix, $H$ the dimensionless Hamiltonian and
the brackets denote the commutator. In the basis of the eigenstates $%
\left\vert l,m\right\rangle $ the coupling matrix is identical to that
derived for Eq. (\ref{NRWA_SE}). Again, a maximum value $C_{max}$ was used,
and the computation was carried out independently for each of the four
conserved parity combinations of $l,m$. Due to the increased order of the
ODE, in several cases the simulations used the RWA coupling instead. In this
case and due to the conservation of $C=l-m$, the calculation was done using
independent "chains" of equal $C$ values, up to $C_{max}$ and according to
the different parity choices for $l,m$. In all simulations the final time of
the simulation was taken to be large enough, so that a clear separation was
achieved between the population around the target state and that left in the
lower $l$ states. This is especially important for values of $P_{1}P_{2}$
near the threshold $1/2$, where such separation is hard to achieve. The
numerical uncertainty in Figs. \ref{Fig3},\ref{Fig4} is smaller than the
marker sizes.



\begin{thebibliography}{99}
\bibitem{SideNote1} Note, that a term of $-1/2$ was omitted from the
function $F\left(L_{z}/L\right)$ in Eq. (7) of \cite{OurPaper} as it carried
no classical meaning.
\end{thebibliography}
\end{document}